\documentclass[twocolumn,superscriptaddress]{revtex4-1}
\usepackage{graphicx}
\usepackage{amssymb}
\usepackage{amsmath}
\usepackage{epsfig}
\usepackage{color}
\usepackage{natbib}
\usepackage{mathtools}
\usepackage[colorlinks,linkcolor=blue,anchorcolor=blue,citecolor=blue,urlcolor=blue]{hyperref}
\usepackage[left]{lineno}
\usepackage{blindtext}
\usepackage{color, soul}
\usepackage{setspace}
\usepackage{fancyhdr}

\begin{document}

\title{Quantifying the performance of multi-pulse quantum sensing}
\author{Yang Dong}
\author{Shao-Chun Zhang}
\author{Hao-Bin Lin}
\author{Xiang-Dong Chen}
\affiliation{{CAS Key Lab of Quantum Information, University of Science and Technology of China, Hefei,
230026, P.R. China}}
\affiliation{{CAS Center For Excellence in Quantum Information and Quantum Physics, University of Science
and Technology of China, Hefei, 230026, P.R. China}}
\author{Wei Zhu}
\author{Guan-Zhong Wang}
\affiliation{{Hefei National Laboratory for Physical Science
at Microscale, and Department of Physics, University of Science and Technology of China, Hefei, Anhui 230026, P. R. China}}
\author{Guang-Can Guo}
\author{Fang-Wen Sun}
\email{fwsun@ustc.edu.cn}
\affiliation{{CAS Key Lab of Quantum Information, University of Science and Technology of China, Hefei,
230026, P.R. China}}
\affiliation{{CAS Center For Excellence in Quantum Information and Quantum Physics, University of Science
and Technology of China, Hefei, 230026, P.R. China}}
\date{\today}

\begin{abstract}
{The quality of a quantum operation determines the performance of quantum information processing, such as the sensitivity of quantum sensing. Different from the fidelity of quantum operation in quantum computation, we present an effective function to evaluate the performance and diagnose the imperfection of operations in multi-pulse based quantum sensing. The evaluation function directly links the realistic sensitivity with intrinsic sensitivity in a simple way. Moreover, guided by this evaluation function, we optimize a composite pulse sequence for high sensitivity nitrogen-vacancy-center based magnetometry against spectrum inhomogeneities and control errors to improve the signal-to-noise ratio of nanoscale nuclear magnetic resonance by $1$ order of magnitude. It marks an important step towards quantitative quantum sensing with imperfect quantum control in practical applications.}
\end{abstract}
\maketitle

Benefitting from quantum superposition \cite{RevModPhys.89.041003} and high performance quantum operation \cite{huang2019fidelity}, quantum information processing (QIP) has
provided great advantages over its classical counterpart in quantum computation \cite{arute2019quantum}, simulation \cite{keesling2019quantum}, secure communication \cite{lucamarini2018overcoming}, and sensing \cite{RevModPhys.89.035002}. Many fundamental concepts have been developed to enhance the performance of quantum operation, such as dynamical decoupling (DD) protocols \cite{kotler2011single,du2009preserving,dong2016reviving}, quantum memory \cite{lovchinsky2016nuclear}, and quantum error correction \cite{PhysRevLett.116.230502}, as well as  quantum feedback control \cite{hirose2016coherent}. In quantum computation, it is challenged by running fault-tolerant operation in an exponentially large computational space \cite{arute2019quantum} to reduce final computation errors. While, the key evaluation of the quantum control for practical quantum sensing is the sensitivity or signal-to-noise ratio (SNR) for a given integral time \cite{kurizki2015quantum,RevModPhys.89.035002}, which distinguishes quantum sensing from other branches of QIP \cite{RevModPhys.89.041003}.

In quantum sensing, a quantum sensor is expected to provide a strong response to wanted signals \cite{RevModPhys.88.021002}. While, it should be minimally affected by imperfect operation under realistic decoherence environment.
In quantum sensing schema, the unknown wanted signal interacts with the quantum sensor and its physical quantity can be stored in the relative phase of the sensor qubit \cite{kotler2011single,dong2016reviving,lovchinsky2016nuclear,PhysRevLett.116.230502,bonato2016optimized}. Once the sensing equipment is fixed, the realistic detection sensitivity is then determined by different operations and limited by the intrinsic sensitivity of system. Currently, the heavily used multi-pulse operation based on DD \cite{RevModPhys.88.041001}, such as Car-Purcell-Meiboom-Gill (CPMG), XY4, and Knill DD (KDD), is the most important and flexible method \cite{RevModPhys.89.041003,kotler2011single}, which can provide ultra-sensitive detections \cite{boss2017quantum,schmitt2017submillihertz} in practical quantum sensing. However, imperfect operations will give significant impact on the performance of sensing protocol and even create a spurious signal \cite{PhysRevX.5.021009,PhysRevA.99.012110,shu2017unambiguous}.

In this Letter, we present an effective function, $F_{QS}$, to evaluate the performance of the operations and diagnose the source of imperfections in DD sequences based multi-pulse quantum sensing. The value of $F_{QS}$ demonstrates the quality of the quantum operation, which links the realistic sensitivity ($\eta_r$) with intrinsic sensitivity ($\eta _{in}$) through a simple relationship
\begin{equation}
{{\eta _r}\text{  =  }}\frac{\eta _{in}}{F_{QS}} \text{.}
  \label{Eq3}
\end{equation}
Furthermore, guided by this evaluation function, we can parameterize the phase and duration of control pulse and optimize them with experimental testing to meet high fidelity and robustness requirements for quantum sensing with intrinsic sensitivity \cite{xu2018room}. As a promising candidate of relevance to many applications in quantum sensing \cite{rondin2014magnetometry,zhang2019thermal,zheng2019zero,li2018enhancing,PhysRevApplied.11.064024,bucher2019quantum,PhysRevApplied.12.044039,zhang2019thermal,xia2019nanometer}, we experimentally demonstrated our method by enhancing the sensitivity and homogeneity in nitrogen-vacancy (NV) center-based nanoscale nuclear magnetic resonance (NMR) detection. Such a method can be directly incorporated to arbitrary DD sequences and is scalable for many other quantum sensing systems.

\begin{figure}[tbp]
\centering
\textsf{\includegraphics[width=8.6cm]{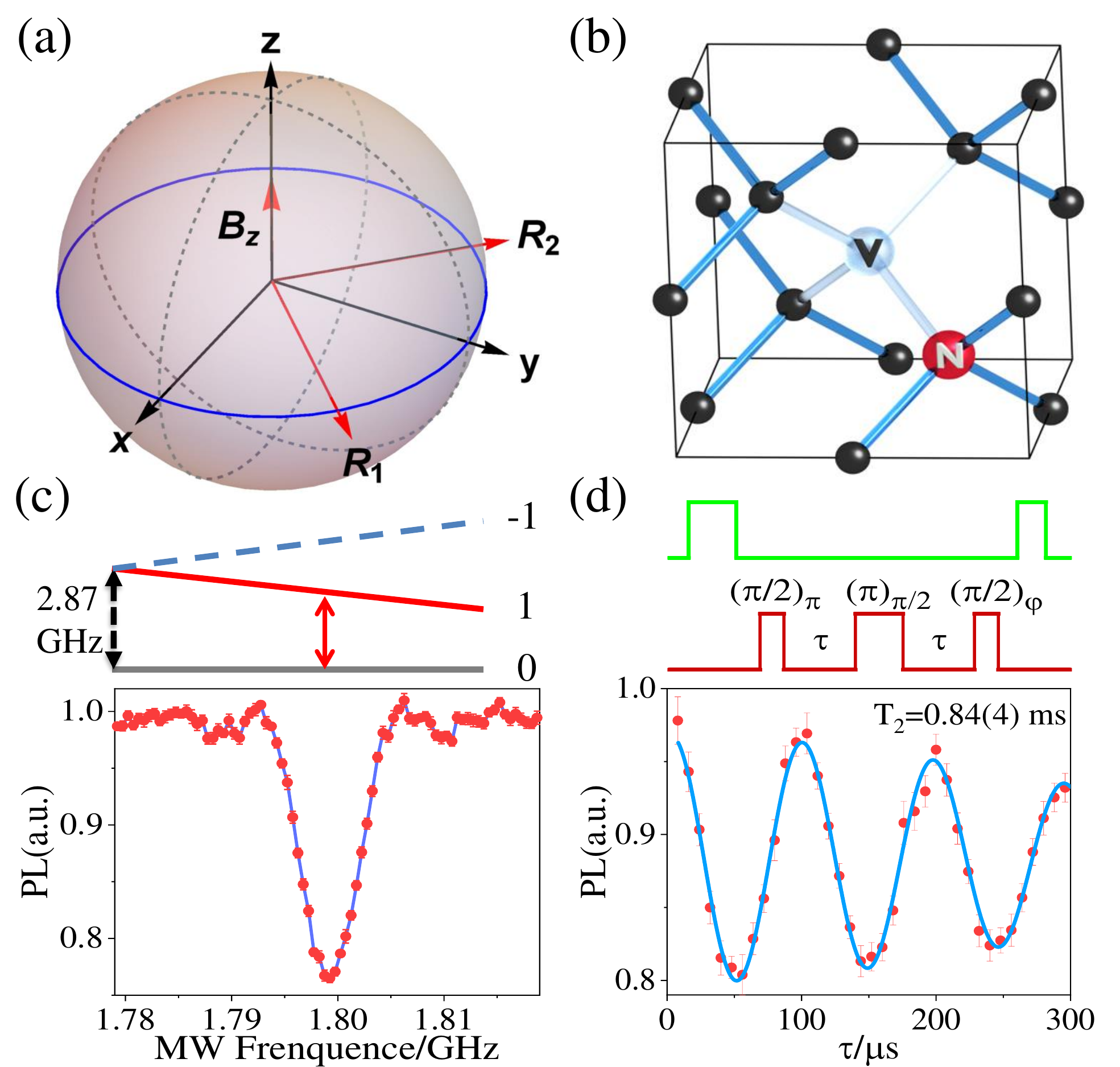}}
\caption{(a) Bloch sphere and the detected signal along $z$ axis. The axis of coherent $\pi$ rotation lies in the equatorial plane, denoting with the blue cure operation. (b) Schematic physical structure of the NV center.
(c) Energy levels of NV center. The ground state of NV center is an electron spin triplet state, with three sublevels ${\left| {m_s} = 0 \right\rangle } $ and ${\left| {m_s} = \pm1 \right\rangle } $. A pulsed ODMR spectrum of ${\left| {m_s} = 0 \right\rangle } \leftrightarrow {\left| {m_s} = 1 \right\rangle }$ with external magnetic field of $382$ Gauss along the NV symmetry axis. (d) Pulse sequence and result of the spin-echo experiment for NV center. The last $\pi /2$ pulse increases linearly with time in the rotating frame.
}
\label{fig1}
\end{figure}

We treat the quantum sensing interaction in the general form $H(t) = \frac{1}{2}{\hat \sigma _z}\hat B(t)$ between a sensor qubit and environment \cite{dong2016reviving}. Here, ${\hat \sigma _z}$ is the Pauli operator of the sensor qubit, and $\hat B(t)$ is an operator that includes the detected signal that oscillates at a particular frequency as well as the presence of noisy environmental fluctuations. So to improve the sensing sensitivity, usually DD operations are experimentally applied to rotate the sensor qubit around the axis in $x-y$ plane of the Bloch sphere with operators formed by the basis of $\left\{ {{{\hat \sigma }_x},{{\hat \sigma }_y}} \right\}$, as shown in Fig. \ref{fig1}(a). In the DD process, $\pi$ gate is a key operation. Hence with a $\pi$ gate, the total spin Hamiltonian is $H(t) = f(t){\sigma _z}\hat B(t)$, where $f(t)$ is the DD modulation function jumping between $ + 1$ and $ - 1$ each time the senor spin is flipped. For periodic DD control, $f(t) = f(t + T)$ with the period $T$, once the modulation frequency (${\omega _{\text{m}}} = \frac{{2\pi }}{T}$) matches with the signals to be measured, the quantum lock-in amplifier is carried out to separate the signal from its noise environment. The performance of this method is determined by the flip-flopping effect of DD unit operation. Hence, we define a new evaluation function for this scheme as:
\begin{equation}
{{F}_{QS}} = \sum\limits_{i = 1,2} {\mathrm{Tr}\left( {\xi {R_{i}}} \right) \text{,}}
  \label{Eq2}
\end{equation}
where $\xi$ is a realistic operation process for DD and ${R_{i}}$ $(i = 1,2)$ are coherent rotations with mutually orthogonal axes lying in the equatorial plane of the Bloch sphere, which can be represented by ${\sigma _i}$ $(i = x,y)$, as shown in Fig. \ref{fig1}(a). The physical meaning of $F_{QS}$ is the successful probability of flip-flopping a sensor qubit. If the ideal channel is unitary, the evaluation reduces to
${{F}_{QS}} = \frac{1}{2} - \frac{{{\mathrm{Tr}}\left( {{\sigma _z}\xi {\sigma _z}{\xi ^\dag }} \right)}}{4}$.
When $\xi {\sigma _z}{\xi ^\dag } =  - {\sigma _z}$, which means flipping sensor qubit round arbitrary axis lying in the equatorial
plane of Bloch sphere completely, the value of the evaluation function reaches the maximum ${F}_{QS}=1$. Otherwise, the DD method will be failed and create undesired signal. This evaluation function is different from the fidelity of quantum operation in quantum computation, which is defined as \cite{PhysRevA.71.062310,nielsen2002simple}
\begin{equation}
{F_{QC}} = \frac{1}{2}\mathrm{Tr}(\xi {U^\dag }) \text{.}
  \label{Eq3}
\end{equation}
$F_{QC}$ evaluates the overlap between realistic operation $\xi$ and target fixed operation $U$. As we will see, the value of $F_{QS}$ has a direct relationship with experimental results in multi-pulse quantum sensing. Moreover, it is more suitable than $F_{QC}$ as a guideline to design multi-pulse sequence for high sensitivity quantum sensing in realistic environments.

A promising candidate of room-temperature quantum sensing qubit is a single NV center in diamond as shown in Fig. \ref{fig1}(b), which will be our experimental testing platform to benchmark the new definition for multi-pulse quantum sensing. To address NV center, we built a confocal microscope with a dry objective lens (N.A. $= 0.9$) at room-temperature \cite{dong2018non}. The driving microwave (MW) was generated by an arbitrary waveform generator (Agilent M8190a) and amplified by microwave amplifier. The NV center studied in this work was formed during the growth of a single-crystal diamond \cite{xu2018room,zhang2019thermal}. Its electronic ground state is a spin triplet ($S = 1$), with an energy splitting $D$ of $2.87$ GHz between states ${m_s} = 0$ and ${m_s} =  \pm 1$. For all experiments in this work, a bias magnetic field aligned with the NV axis split the degenerate spin states, allowing selective addressing of the transition, which was represented with the qubit states $\left| {\text{0}} \right\rangle $ and $\left| 1 \right\rangle $. The spin-dependent photon luminescence (PL) enabled the implementation of optically detected magnetic resonance (ODMR) technique to detect the spin state as shown in Fig. \ref{fig1}(c). The undesired couplings between the qubit and the surrounding $^{13}C$ nuclear spins led to the dephasing effect with $T_2^*=5.6(1)$. This effect can be suppressed by two orders of magnitude by DD method with $T_{2}=0.84(4)$ ms, as shown in Fig. \ref{fig1}(d).

\begin{figure}[thbp]
\centering
\includegraphics[width=8.6cm]{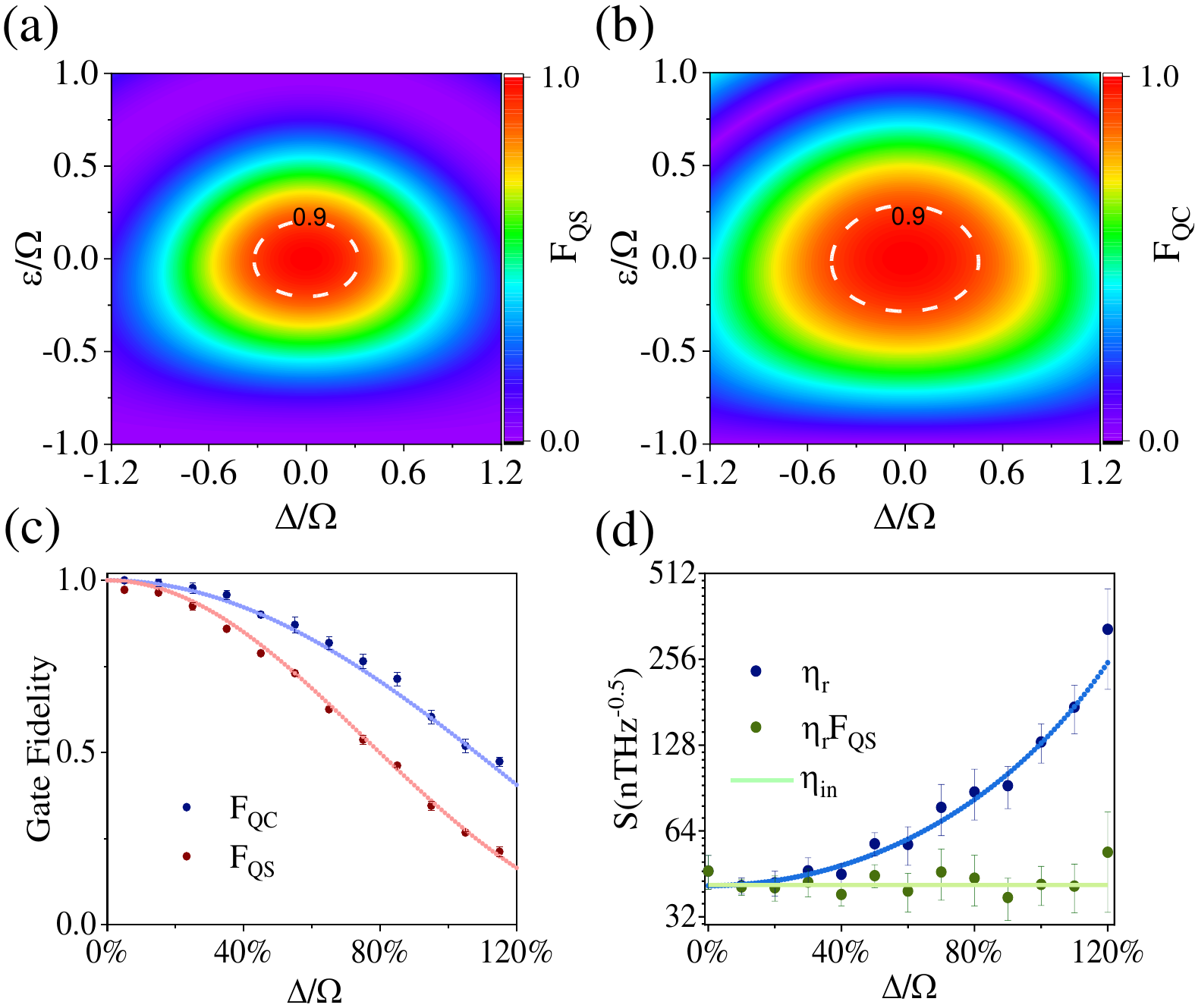}
\caption{(a)-(b) Simulated results of $\pi$ gate with a rectangular pulse according to $F_{QS}$ and $F_{QC}$ with a range of detuning ($\Delta$) and control amplitude ($\varepsilon$), scaled by Rabi frequency ($\Omega$) respectively. (c) Measured results and simulated fidelity of $\pi$ pulse with different detunings at $\varepsilon=0$. (d) Measured magnetometry realistic sensitivities ($\eta_r$) of spin-echo protocol for a single NV center using rectangular $\pi$ pulse (deep blue dots). The deep green dots show $\eta_r F_{QS} $, while the bright green straight line shows the intrinsic sensitivity ${\eta _{in}}{\text{ = }} 41\text{nT}  /\sqrt{\text{Hz}}$ \cite{SM}.}
\label{fig2}
\end{figure}

At first, we systematically investigated the performance of general rectangular $\pi$ gate against the noise from the quasistatic fluctuation of the magnetic field, which is simulated by the detuning frequency ($\Delta$) from the on-resonance frequency and amplitude fluctuation ($\varepsilon$) of MW. Here $\pi$ gate will not be a perfect flip operation anymore when $\Delta\neq0$ or $\varepsilon\neq0$. Typical simulation results of $F_{QS}$ and $F_{QC}$ from Eq. (\ref{Eq2}) and Eq. (\ref{Eq3}) are shown in Fig. \ref{fig2}(a) and (b), respectively. We can find that the fidelities defined in both Eq. (\ref{Eq2}) and Eq. (\ref{Eq3}) give similar relationship with various frequency detunings and amplitude fluctuations of MW except for the detailed numerical values. The value of $F_{QS}$ drops a little more quickly than $F_{QC}$ as shown in Fig. \ref{fig2}(c), which means that the flip effect of rectangular $\pi$ pulse is not robust with detunings. And the physical reason is that the practice $\pi$ gate with detuning will rotate around some other axis, which lies in $x-z$ plan of Bloch sphere. Hence, the elements of common rectangular $\pi$ gate will transfer in the operation space of ${\{ I},{\sigma _x},{\sigma _z}{\text{\} }}$, which can be witnessed by QPT in experiment (see the Supplemental Material for details \cite{SM}). As shown in Fig. \ref{fig2}(c), the values of $F_{QS}$ and $F_{QC}$ for this gate decrease monotonously with the detuning when $\varepsilon=0$.

Furthermore, in the experimental testing, we performed a synchronization ac magnetic field sensing protocol by employing DD method \cite{dong2016reviving,PhysRevX.5.041001,RevModPhys.89.035002,RevModPhys.88.041001}. The typical spin-echo magnetometry \cite{RevModPhys.89.035002,xu2018room} with rectangular pulse was performed by changing the amplitude of ac magnetic field and sensing time \cite{SM}. For a fixed sensing time, the accumulated phase has a linear relationship with the amplitude of ac magnetic field and the best sensitivity \cite{RevModPhys.89.035002,li2018enhancing,xu2018room} of a magnetic field measurement is given by ${\eta _{r}} =\frac{\sigma }{{dS/d{B_{un}}}}\sqrt \tau  $, where the standard deviation ($\sigma $) of the sensing signal is compared to the response of the system $dS$ in a changing magnetic field $d{B_{un}}$. $\tau$ is the optimal sensing time. The final detection sensitivity for rectangular $\pi$ gate are shown in Fig. \ref{fig2}(d) with various detunings. We find that the realistic sensitivity (${\eta _r}$) with different detunings can be normalized to $\eta_{in}$ by the new evaluation function in a simple way ${\eta _{in}}\text{  =  }{\eta _r}F_{QS}$. Here ${\eta _{in}}$ was separately calculated from the experimental data and the decoherence time $T_2$ \cite{SM,PhysRevLett.115.190801}, as shown in Fig. \ref{fig2}(d) with a bright green line. Therefore, $F_{QS}$ has shown the capability of evaluating the performance of the operation in quantum sensing based on the sensitivity.

\begin{figure}[tbp]
\centering
\textsf{\includegraphics[width=8.6cm]{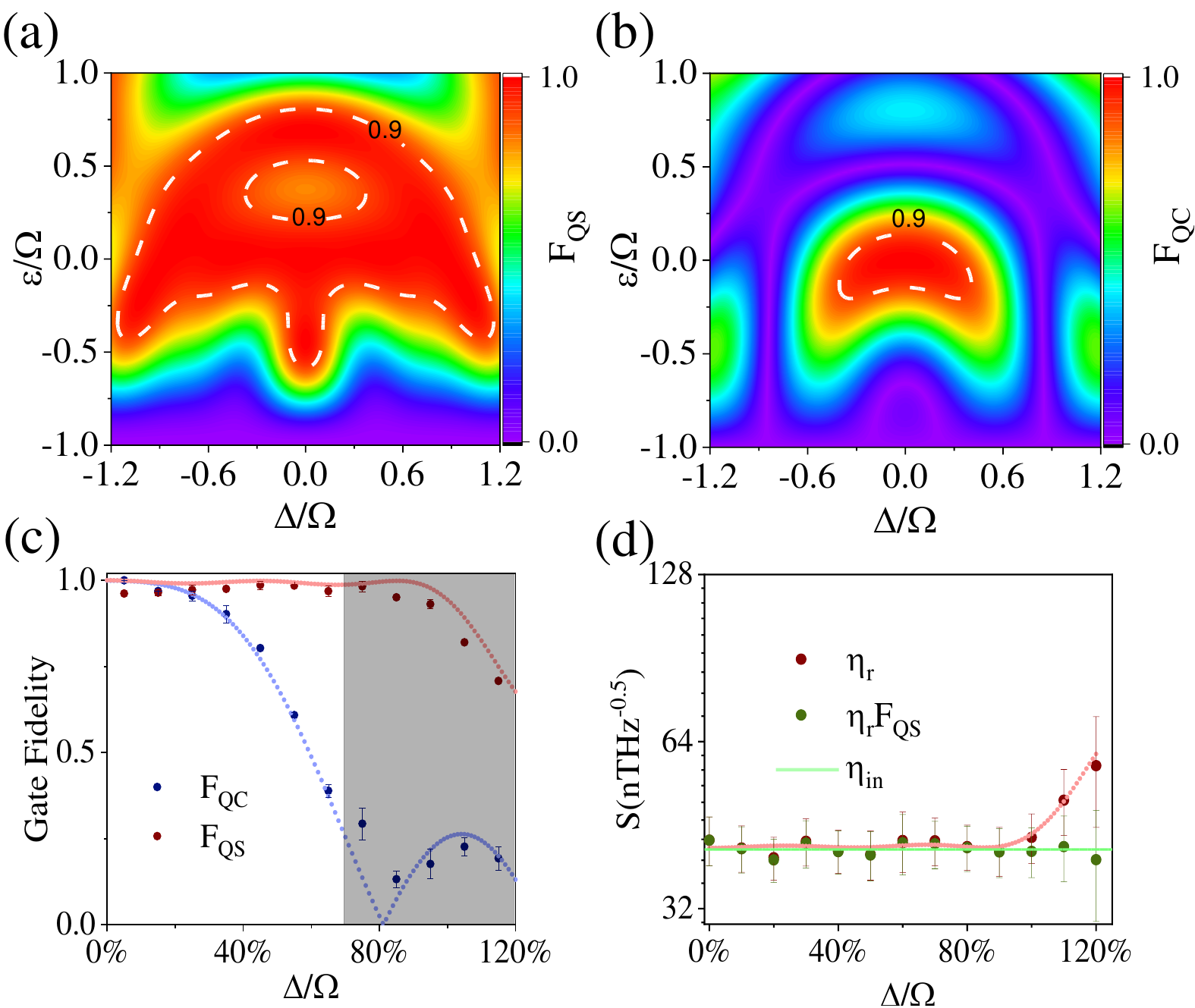}}
\caption{{(a)-(b) Simulated results of $\pi$ gate with optimized composite pulses according to $F_{QS}$ and $F_{QC}$ with a range of detuning and control amplitude. (c)  Measured results and simulated fidelity of $\pi$ pulse with different detunings at $\varepsilon=0$.
The non-monotonic relationship for $F_{QC}$ dependence on detunings are highlighted with a shadow region.
(d) Measured magnetometry realistic sensitivities of spin-echo protocol using composite $\pi$ pulse (red dots). The deep green dots show $\eta_r F_{QS}$, while the bright green straight line shows the intrinsic sensitivity ($\eta_{in}$).}}
\label{fig3}
\end{figure}

Since $F_{QS}$ can give a quantitative description of realistic detection sensitivity in a simple and direct way, we can employ it as a guideline to design a quantum optimal control to realize robust quantum sensing with intrinsic sensitivity. For magnetometry sensing based on NV ensembles, high densities increase the fluorescence yield to improve the sensitivity at the price of increased inhomogeneous broadening due to the dipolar magnetic interaction with paramagnetic impurities in the host crystal. And in principle, the inhomogeneous broadening effect on the multi-pulse quantum sensing can be simulated with a finite detuning \cite{SM}. After taking this major effect into consideration, we designed an effective $5$-piece composite-$\pi$-pulse under the guideline of Eq. (\ref{Eq2}) with gradient ascent algorithm \cite{xu2018room}, where $R(\pi ) = {\left( {0.5\pi } \right)_x}{\left( {1.12\pi } \right)_y}{\left( {0.44\pi } \right)_{ - y}}{\left( {1.12\pi } \right)_y}{\left( {0.5{\pi }} \right)_x}$. The duration time of composite pulse is the same as the order of magnitude as the rectangular $\pi$ pulse and increases much more slowly than the results \cite{xu2018room} based on Eq. (\ref{Eq3}).

The final results with such control composite-pulses are shown in Fig. \ref{fig3}(a)-(b).
Obviously, $F_{QS}$ give a significant different evaluation compared to that obtained by $F_{QC}$. This optimized composite pulse can be recognized as a robust $\pi$ gate to enhance the coherent qubit-flip effect over a large range of detuning and control amplitude fluctuation by comparing Fig. \ref{fig2}(a) and Fig. \ref{fig3}(a). However, from the view of quantum computation based on $F_{QC}$, the fidelity of this composite pulse decreases sharply with detunings or control amplitude fluctuations and gives a negative evaluation by comparing Fig. \ref{fig3}(b) with Fig. \ref{fig2}(b). Then we experimentally applied this optimized composite pulse sequence on a single NV center. As shown in Fig. \ref{fig3}(c), without MW amplitude fluctuation, the fidelity ($F_{QC}$) of this composite operation even degenerates into $0$ at some detuning (at $80\%$ of the resonant Rabi frequency). However, the value of $F_{QS}$ did not change much and can keep at a high level with the detuning as much as $ {\text{100\% }}$ of the Rabi frequency. Therefore, $F_{QS}$ can be used to quantify the perfection of flip operation. By taking snapshots of the process for different detunings, we monitored how the process matrix element transfers from the ${\sigma _x}$ operation to the space of $\{ {\sigma _x},{\sigma _y}\} $ \cite{SM}. So the fidelity ($F_{QC}$) of this composite pulse, which presents the overlap between practice and ideal operations, decreases with the increasing detuning. But the flip effect of this composite pulse will remain unchanged for a larger detuning range. At last, when the detuning is larger than $100\%$ of Rabi frequency, the practice composite operation exceeds the space of $\{ {\sigma _x},{\sigma _y}\} $ and both values of the two evaluation functions drop finally as shown in Fig. \ref{fig3}(c). After applying this composite pulse to the spin-echo magnetometry of NV center, the sensitivity can keep constant up to $100\%$ detuning, agreeing with theoretical prediction as shown in Fig. \ref{fig3}(d). The realistic sensitivity (${\eta _r}$) with various detunings can be also normalized by Eq. (\ref{Eq3}). When the detuning increases to $115\%$ of the resonant Rabi frequency, we experimentally improved the sensitivity by a factor of $4$, comparing to the control with normal rectangular pulse.

\begin{figure}[tbp]
\centering
\textsf{\includegraphics[width=8.6cm]{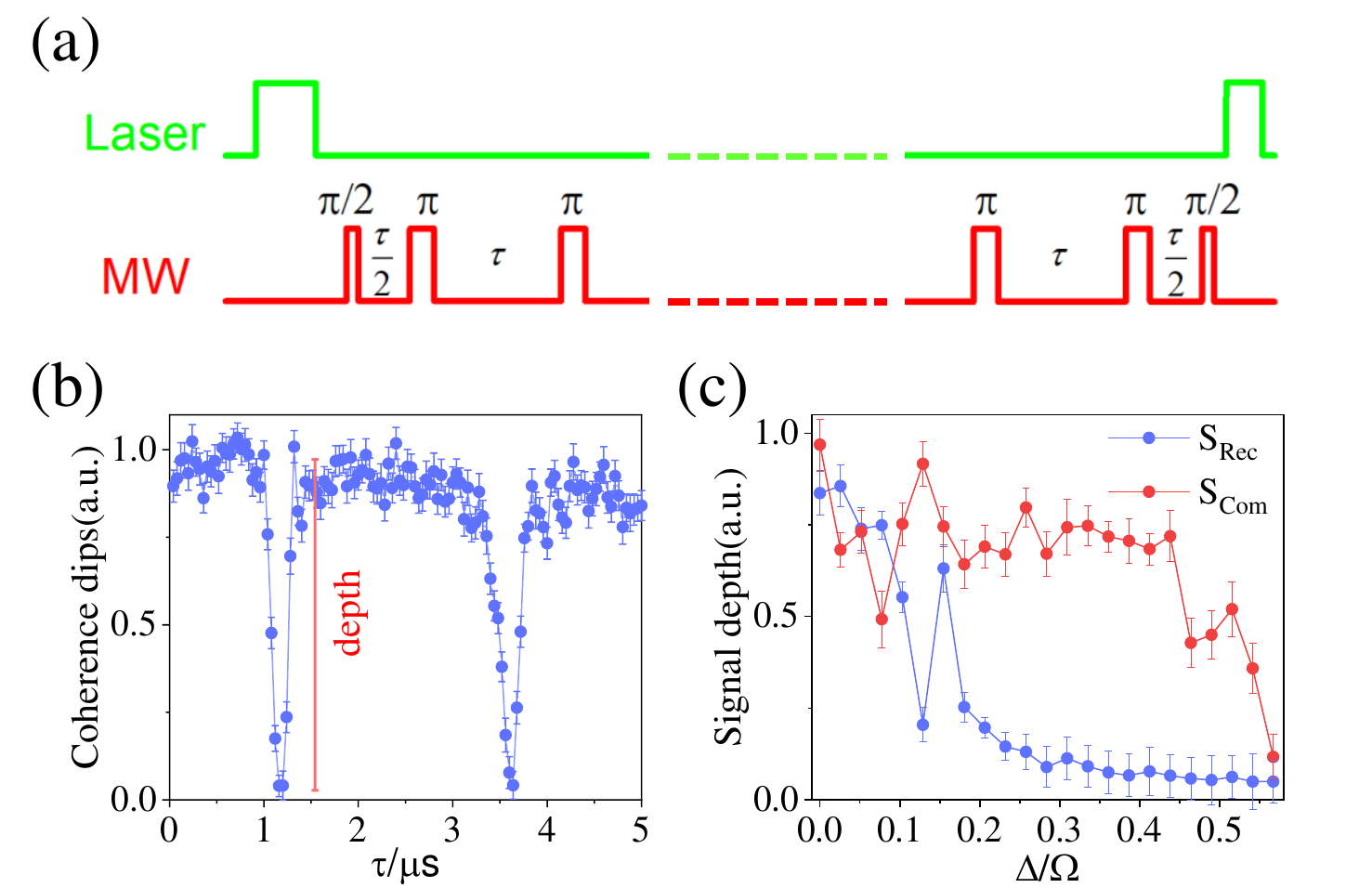}}
\caption{{(a) CPMG sequence for nanoscale NMR signal detection with NV center. The number of $\pi$ pulses was $16$. (b) Detection of NMR signal of $^{13}C$ with scanned pulse spacing. The depth of collapses can be extracted by fitting with Gaussian lineshape.
(c) Measured depth of nanoscale NMR signal with different detunings with rectangular (blue dots) and optimized composite (red dots) pulses, respectively.}}
\label{fig4}
\end{figure}

Having shown the robustness of the flip effect with the optimized composite control pulses using NV center, we further employed it for the detection of nanoscale NMR \cite{SM}. NMR spectroscopy is an important test case for our method because long time duration of composite always limits its application in experiment \cite{xu2018room,RevModPhys.88.041001,PhysRevLett.112.050503}. Without detuning, we detected the NMR signal
of $^{13}{C}$ nuclei located in close proximity to the NV center with multi-pulse quantum sensing sequences as shown in Fig. \ref{fig4}(a). The experimental signal of multi-pulse detection is shown in Fig. \ref{fig4}(b). When the time of the multi-pulse separation meets the resonance condition \cite{PhysRevLett.109.137602} ${\tau _k} \approx \frac{{(2{\text{k - 1}})\pi }}{{2{\omega _L}}}$, collapses will appear in coherence trace. The collapses correspond to the overlapping signals of multiple nuclear spins in the spin bath, whose product leads to coherence disappear \cite{yang2016quantum,PhysRevApplied.6.024019}. The SNR of NMR is an important index and can be directly characterized by the depth of collapses \cite{PhysRevLett.113.030803,barry2019sensitivity}. With the optimized composite pulses, the SNR remained constant over a large recorded parameter range. While for the case with rectangular pulse, it dropped off by $1$ order of magnitude as the detuning becomes large. Hence, the multi-pulse quantum sensing with the concatenation of composite pulse sequence can give a significant improvement in the SNR for NMR compared with traditional case.

In conclusion, the concept of an effective evaluation function for DD-based quantum sensing has been implemented in theory and experiment. Under the guideline of the definition, we optimized the core $\pi$ gate operation by quantum optimal control algorithms. We have showed how a high sensitivity in an NV-center magnetometry can be maintained over a wide range of control parameters by reducing the effects of inhomogeneous broadening. And the value of evaluation function also have simple relationship between the realistic and intrinsic sensitivity. Furthermore, by nesting composite sequences into multi-pulse quantum sensing protocol, we improved the SNR of nanoscale NMR by $1$ order of magnitude for same integral time at room-temperature. Beyond the current experiment results, the advantage of the evaluation function is robust and incorporated directly with arbitrary DD sequences. And when the multiple sensor qubits are entangled with specific MW circuit \cite{PhysRevB.100.214103}, the optimized composite pulse based DD can protect \cite{dong2016reviving} the multi-qubit quantum metrology in noisy environment to achieve the Heisenberg limit for practical applications.

\section*{Acknowledgment}
This work is supported by The National Key Research and Development Program of China (No. 2017YFA0304504), the National Natural Science
Foundation of China (Nos. 91536219, and 91850102), Anhui Initiative in Quantum Information
Technologies (No. AHY130000), the Science Challenge Project (Grant
No. TZ2018003).

\clearpage
\newpage
\leftline{\textbf{Supplemental Material}}
\section{Operation and Signal Interpretation}
\begin{figure}[tbp]
\centering
\textsf{\includegraphics[width=8cm]{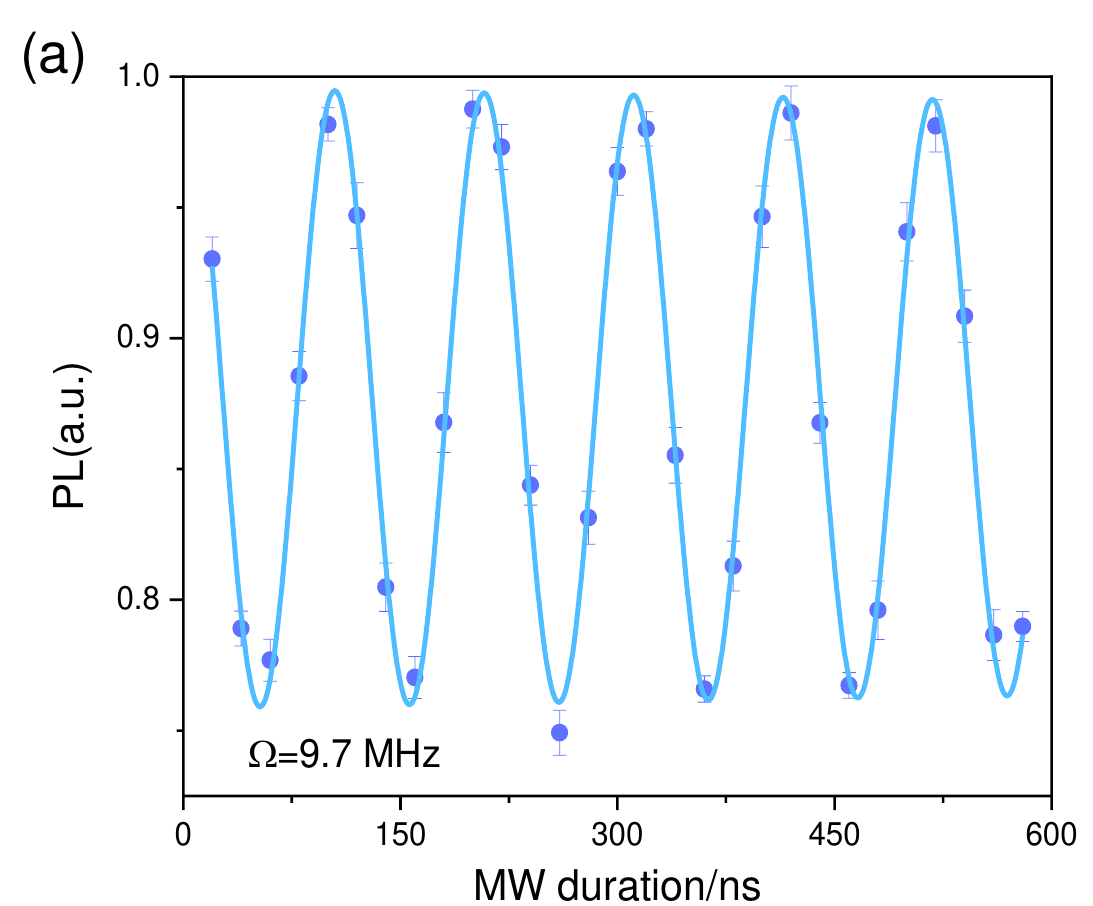}}
\caption{{ Results of the nutation experiment for the electron spin of NV center. The decay time of the nutation is ${T_{{\text{1}}\rho }}{\text{ = 0}}{\text{.14(1)}}$ ms. The signal contrast $C=0.24(1)$. The solid blue line in the panel is a fitting to the experimental results.}}
\label{Sfig1}
\end{figure}

The normalization of the coherent MW operation for ODMR, Ramsey and spin-echo magnetometry signals in the main text were carried out by performing an electron spin nutation experiment as shown in Fig. \ref{Sfig1}. And the Rabi frequency was $9.7$ MHz by fitting the experimental data. From the decay constant of Rabi oscillation signal, the fluctuation ($\varepsilon$) of MW amplitude was much smaller than intrinsic inhomogeneous broadening of NV center. Therefore we can set $\varepsilon=0$ in the experimental demonstration.

In the experiment, the observed fluorescence signal, which is related to the
population distributions between ${m_s} = 0 $ and ${m_s} = +1$ states of NV
center, can be converted to the successful probability of quantum coherent
operation by linear transformation. Specially, we firstly initialized the
system into ${m_s} = 0 $ state by a laser pulse and then changed it into ${m_s} = +1$ state by MW operation and measured their photon counts (denoted
by $I_{\max}$, $I_{\min}$). In order to beat the fluctuation of photon
counting, we repeated the experimental cycle at least ${10^6}$ times. So the
relative population of the ${m_s} = 0 $ state for an unknown state can be
expressed as
\begin{equation}
{P_0} = \frac{{I - {I_{\min }}}}{{{I_{\max }} - {I_{\min }}}} \text{,}
 \label{SEq1}
\end{equation}%
where $I$ is the measured photon count under same experimental condition.
Therefore, the quantum state of NV center electron spin can be determined from the fluorescence intensity of NV center \cite{Sdong2018non}.

\section{Quantum process tomography}
To analyze the performance of the control pulses, we performed quantum process tomography (QPT) and calculated the fidelity of experimental process \cite{Snagata2018universal}. We aimed at characterizing the effect of a process $E$ on an arbitrary input quantum state $\rho $. Any such processes can be characterized in terms of a dynamical matrix $\chi $, via
\begin{equation}
 E(\rho ) = \sum\limits_{i,j = 1}^{\text{4}} {{\chi _{i,j}}} {A_i}\rho A_j^\dag \text{,}
  \label{SEq2}
\end{equation}
where the operators ${A_i}$ formed a complete basis set of operators.

For a qubit system, the tomography matrix $\chi $ has $16$
elements. Four elements follow from the completeness relation $\sum\limits_{i,j = 1}^{\text{4}} {{\chi _{i,j}}} {A_i}A_j^\dag  = E$, and twelve elements have to be
measured. By choosing ${A_i}{= \{ I}$, ${\sigma _x}$, ${\sigma _y}$, ${\sigma _z}{\text{\} }}$, where $I$ is the identity and $\sigma_i$ $(i=x$, $y$, $z)$ are the Pauli spin operators, the task reduces to a rather simple procedure. The qubit is consecutively prepared in each of the four initial states $\left| \psi  \right\rangle  = \left\{ {\left| 0 \right\rangle ,\left| 1 \right\rangle ,\left( {\left| 0 \right\rangle {\text{ + }}\left| 1 \right\rangle } \right)/\sqrt 2 ,\left( {\left| 0 \right\rangle  - i\left| 1 \right\rangle } \right)/\sqrt 2 } \right\}$. NV center can be easily initialized into ${\left| 0 \right\rangle }$ by $532$ nm laser with a few microseconds. The remaining states are coherently transferred from ${\left| 0 \right\rangle }$ by applying $\pi $ and $\pi /2$ gates. For each initial state, the system evolves driven by the control pulse. To determine the final states $E \left( {\left| \psi  \right\rangle \left\langle \psi  \right|} \right)$ by quantum state tomography, we measured the expectation values of the three spin projections ${\sigma _x}$, ${\sigma _y}$, and ${\sigma _z}$ by optical readout \cite{Sdong2018non}. The readout of the projection on each of the
basis states was averaged over ${\text{1}}{{\text{0}}^7}$ shots in experiment. Finally, the complete set of initial states and final states defined the tomography matrix $\chi $ by Eq. (\ref{SEq2}). Effects such as noise and finite sampling of the expectation values can lead to one or more negative eigenvalues of the tomography matrix $\chi $ despite the physically required
property of positivity. And by employing maximum likelihood estimation (MLE) algorithm \cite{Snagata2018universal}, this imperfection can be removed effectively.

\begin{figure*}[tbp]
\centering
\textsf{\includegraphics[width=17cm]{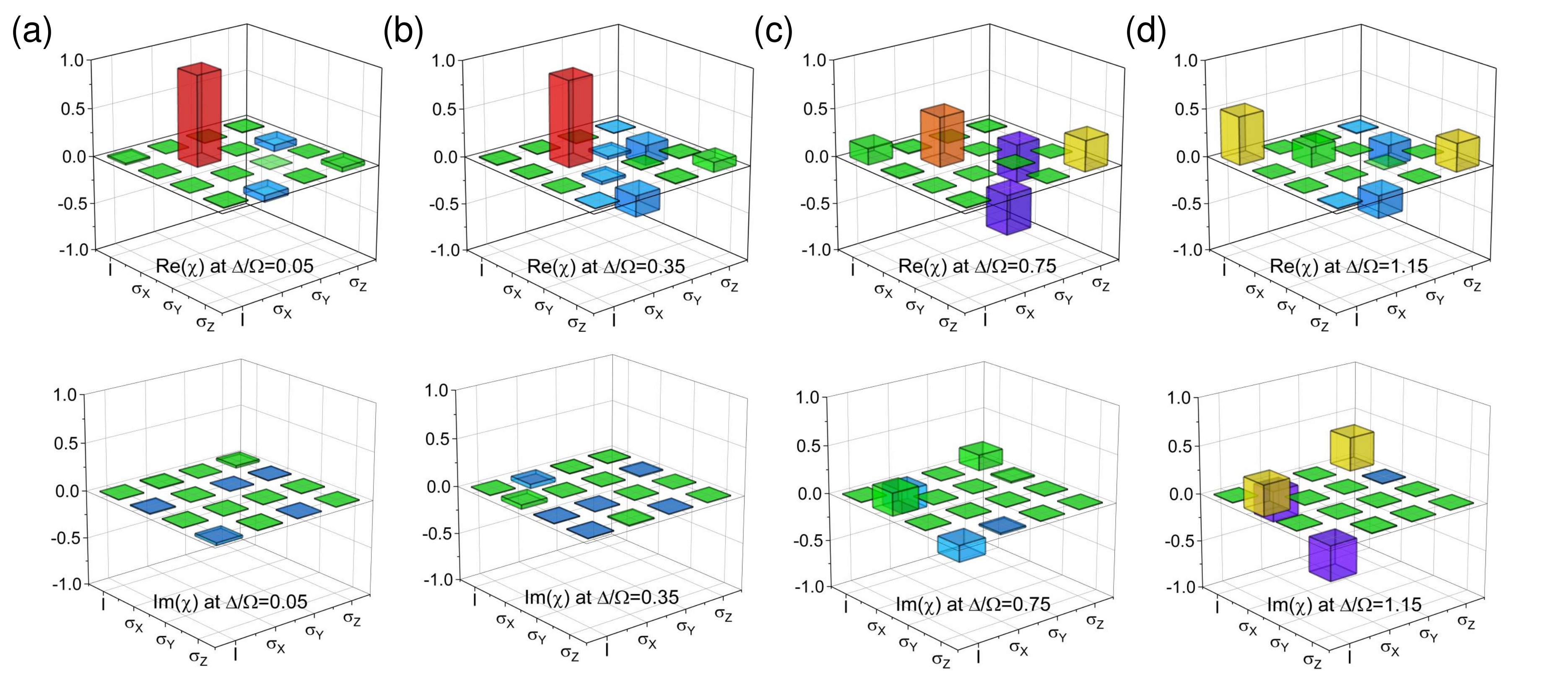}}
\caption{{(a)-(d) Dynamic QPT of a single qubit $\pi$ gate implemented by a common rectangular MW pulse gate by QPT for (a) $5\%$ detuning, (b) $35\%$ detuning, (c) $75\%$ detuning, (d) $115\%$ detuning with $\varepsilon=0$.}}
\label{Sfig2}
\end{figure*}

\begin{figure*}[tbp]
\centering
\textsf{\includegraphics[width=17cm]{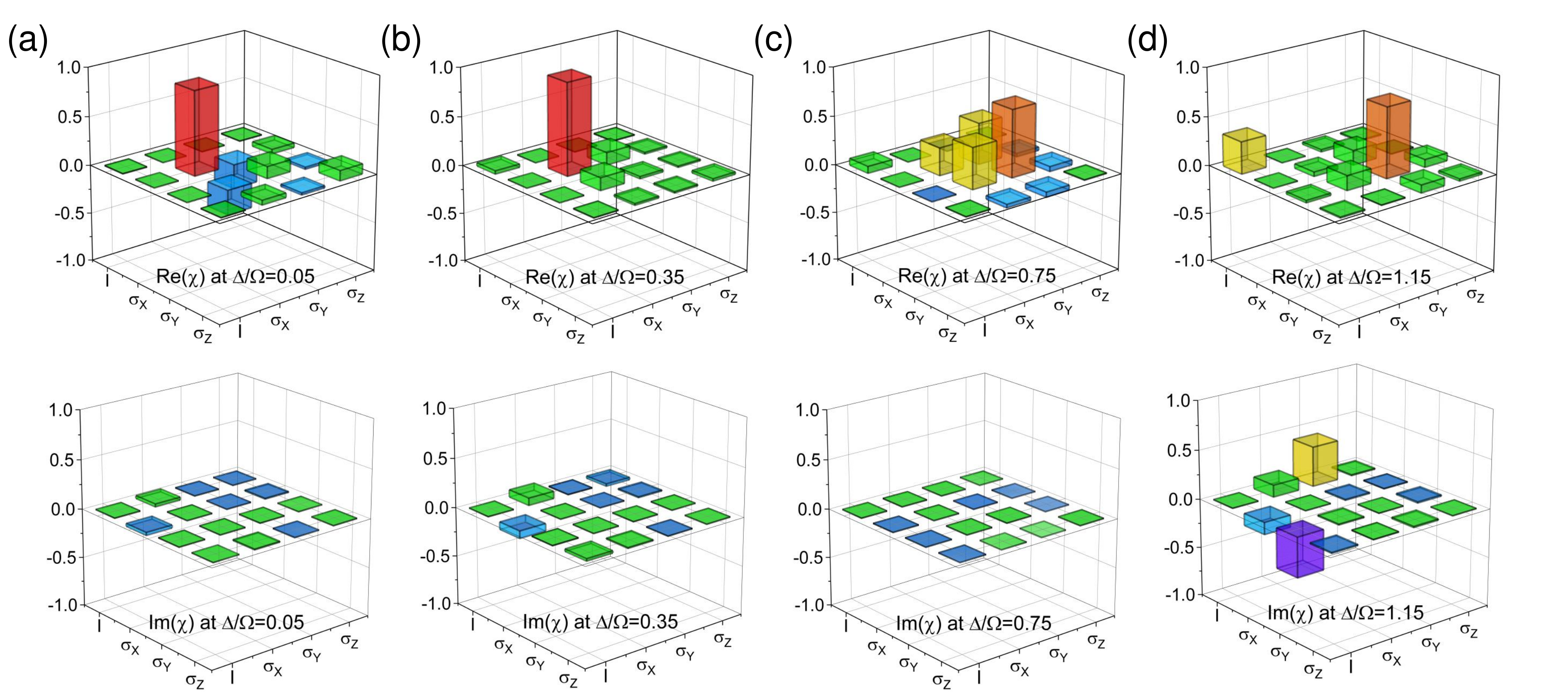}}
\caption{{(a)-(d) Robustness of a single qubit $\pi$ gate implemented by an experimental optimal composite pulse characterized by QST for (a) $5\%$ detuning, (b) $35\%$ detuning, (c) $75\%$ detuning, (d) $115\%$ detuning with $\varepsilon=0$.}}
\label{Sfig3}
\end{figure*}

Typical results of QPT for a $\pi$ gate with common rectangular and optimized composite pulse are shown in Fig. \ref{Sfig2}(a)-(d) and Fig. \ref{Sfig3}(a)-(d), respectively. For the former, with the detuning increasing, the practice $\pi$ gate will rotate around some other axis, which lies in $x-z$ plane of Bloch sphere. Hence, the elements of the common rectangular $\pi$ gate will be transferred in the operation space of ${\{ I},{\sigma _x},{\sigma _z}{\text{\}}}$, as shown in Fig. \ref{Sfig2}(a)-(d). And the fidelity ($F_{QC}$) and the new evaluation function ($F_{QS}$) of this gate will decrease monotonously, as shown in the main text. However, by incorporating $F_{QS}$ constraint into quantum optimal algorithm, the optimized composite pulse has different relationship with detunings as shown in Fig. \ref{Sfig3}(a)-(d), which can be used to improve the performance of DD-based quantum sensing in the main text.

\section{The connection between inhomogeneous broadening and detuning}
In the current experiment, the inhomogeneous broadening of the NV center is smaller than Rabi frequency. So we replaced this effect with detuning effect just like previous work \cite{Swang2012composite,SPhysRevLett.112.050503,SPhysRevLett.115.190801,Srong2015experimental}. There is a clear physical connection between them. Since the Zeeman energy of spin bath under the external field dominates over the hyperfine interaction and the dark spin interaction, and the room-temperature to be considered is much higher than the interaction energy, the thermal spin bath state can be taken as \cite{Sliu2007control}
\begin{equation}
 {\rho ^N} \approx \sum\limits_J {{P_J}} \left| J \right\rangle \left\langle J \right| = \sum\limits_J {{P_J}} \left| J \right\rangle \left\langle J \right| \text{,}
  \label{SEq3}
\end{equation}
where $\left| J \right\rangle  = { \otimes _n}\left| {{j_n}} \right\rangle $ is an eigenstate of spin bath and ${P_J} = \prod\nolimits_n {{p_{{j_n}}}} $ is the probability distribution with ${p_{{j_n}}} \equiv \frac{{{e^{ - {j_{n,a}}{\omega _n}/T}}}}{{\sum\nolimits_{j =  - {J_a}}^{{J_a}} {{e^{ - j{\omega _n}/T}}} }}$ for the population of the single-spin state $\left| {{j_n}} \right\rangle $. So the whole system can be treated as
\begin{equation}
{\rho _s} = {\rho ^e} \otimes {\rho ^N} \approx {\rho ^e} \otimes \sum\limits_J {{P_J}} \left| J \right\rangle \left\langle J \right| = \sum\limits_J {{P_J}} {\rho ^e} \otimes \left| J \right\rangle \left\langle J \right| \text{.}
\label{SEq4}
\end{equation}

In the experiment, we can coherently operate each state composition ${\rho ^e} \otimes \left| J \right\rangle \left\langle J \right|$ with the fidelity greater than or equal to $0.9$ within a fixed detuning region. Because of no quantum coherence among the maximally mixed state of spin bath, the whole fidelity is just an average result and is greater than or equal to $0.9$ with the optimized composite pulse sequence. The inhomogeneous broadening of NV center ensemble can be also described by similar quantum physical model. So the conclusion will still be held.

\section{The spin-echo magnetometry for NV center}

\begin{figure}[bp]
\centering
\textsf{\includegraphics[width=9cm]{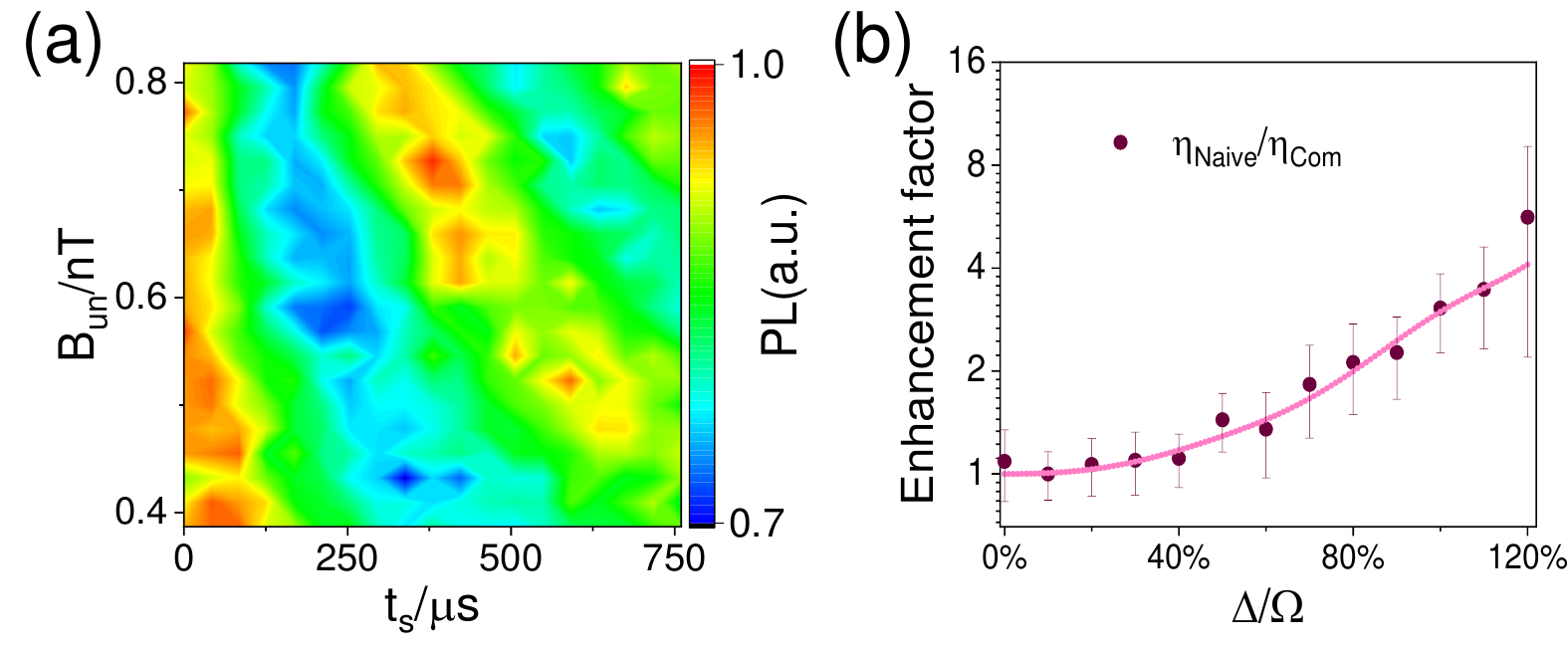}}
\caption{{(a) Results of the spin-echo sensing sequence as a function of sensing time and the amplitude of ac magnetic field. (b) The relative sensitivity enhancement of optimized composite pulse versus rectangular pulse sequence under detuning in spin-echo sensing protocol. }}
\label{Sfig4}
\end{figure}

Following the spin-echo magnetometry protocols \cite{SRevModPhys.89.041003,Sdong2016reviving,Sxu2018room}, which is the simplest DD-based quantum sensing, we measured the projection of an external ac magnetic field $B$ onto the NV symmetry axis by reading out the modulation of the spin echo amplitude. We synchronized the free precession intervals of the spin echo sequence to the half-periods of the external magnetic field. Due to the Zeeman effect, the energy of ${m_s} =+ 1$ get shifted into opposite directions relative to ${m_s} = 0$ component during the first and second free precession periods, respectively.
The resulting phase accumulation $\Delta \varphi $ between ${m_s} = +1$ and ${m_s} = 0$ components of the precessing superposition state is converted to a difference in ${\sigma _z}$ population by a final $\pi/2$ pulse of the sequence and readout by laser finally. For simplicity, we applied $B$ in a square wave with the amplitude ${B_0}$ \cite{Sxu2018room}. The typical result is shown in Fig. \ref{Sfig4}(a). We can define the smallest detectable magnetic field $\delta {B_{\min }}$ as the signal that is the same as the noise, i.e. at a signal-to-noise ratio $\text{SNR} = 1$. The noise in the experiment was limited by the photon shot noise, ${\sqrt {n{t_r}} }$, where ${n} = 4.8 \times {10^4}$ ${\text{counts}}/s$ denoted the average photon number per second and ${t_r}=270$ ns was signal readout time. Hence, the intrinsic sensitivity was ${\eta _{\operatorname{in} }}{\text{  =  }}\frac{{1/\sqrt {n{t_r}} }}{{C{\gamma _{\text{e}}}\sqrt {{T_2}/2} }} = 41\text{nT} /\sqrt{\text{Hz}}$. For comparison, the relative sensitivity enhancement of
composite-pulse versus rectangular pulse sequences is shown in Fig. \ref{Sfig4}(b) under same detuning for spin-echo sensing protocol.

\section{Nanoscale Nuclear Magnetic Resonance Detection}
\begin{figure*}[tbp]
\centering
\textsf{\includegraphics[width=17cm]{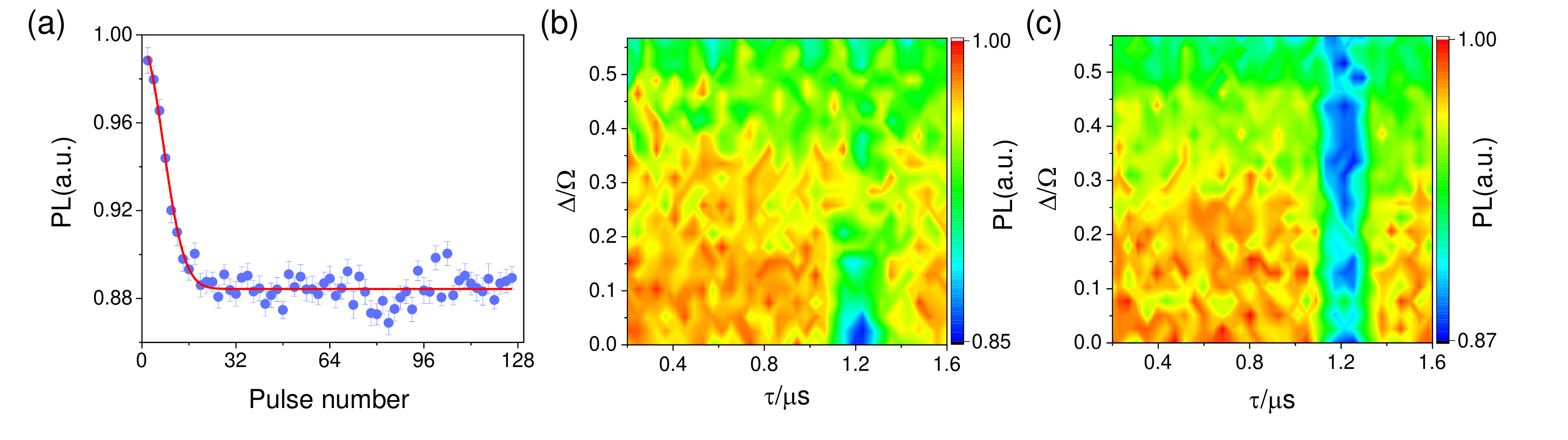}}
\caption{{(a) Sensor coherence dip as a function of the CPMG DD pulse number N for $^{13}C$ spins bath. (b)-(c) NMR spectra obtained using the CPMG sequence for various values of the detuning, with a rectangular or optimized composite $\pi$ pulse under a magnetic field $ B \approx 380$ G. The number of $\pi$ pulses in CPMG sequence was $N = 16$.}}
\label{Sfig5}
\end{figure*}
We can probe $^{13}C$ spin bath of NV center by preparing the electron spin in a superposition, $\left| x \right\rangle  = \left( {\left| 0 \right\rangle  + \left| 1 \right\rangle } \right)/\sqrt 2 $, and applying a dynamical decoupling sequence consisting of $N$ sequential pulses. We used the basic decoupling unit on the electron spin ${\left[ {\tau /2 - \pi  - \tau  - \pi  - \tau /2} \right]^{N/2}}$, in which $\tau $ was a free evolution time. At first, by sweeping the pulse interval with a constant pulse number in the main text, we tuned the electron spin resonance with the target spin bath. At ${\tau _k} \approx \frac{{(2{\text{k - 1}})\pi }}{{2{\omega _L}}}$, a dip will appear at NV center coherence curve in the main text. Then we fixed the pulse interval (${\tau  = 1240}$ ns) and scanned the number of pulse. The results is shown in Fig. \ref{Sfig5}(a). The depth of coherent dip increases with the number of pulse and reaches the maximum once $N \geqslant 16$. We fitted the experimental data with the function $y = a\exp \left( { - \lambda {N^2}} \right) + b$, where $\lambda $ was the effective interaction between NV center and $^{13}C$ spin bath \cite{SPhysRevApplied.6.024019}. Finally, we measured the depth of this signal with different detunings with rectangular and the optimized composite pulse, as shown in Fig. \ref{Sfig5}(b)-(c), respectively. From the comparison, the optimized composite pulse, which is designed by the new evaluation function $F_{QS}$, is robust against with detuning and can be directly applied in high-resolution magnetic resonance spectroscopy based on NV center ensemble \cite{Sglenn2018high}.

The Hamiltonian of interaction between NV center and $^{13}C$ nuclear spin can be written as \cite{SPhysRevB.78.094303}£º
\begin{equation}
\begin{aligned}
H = &DS_z^2 - {\gamma _e}{B_z}{S_z} - \sum\limits_n {{\gamma _N}\vec B \cdot {{{\bf{\vec g}}}_n}(\left| {{S_z}} \right|) \cdot {{\vec I}_n}}  \\
&+ \sum\limits_n {{S_z}{{\hat A}_n} \cdot {{\vec I}_n}}  + \sum\limits_n {\delta {{\hat A}_n}(\left| {{S_z}} \right|) \cdot {{\vec I}_n}} \\
&+ \sum\limits_{n > m} {{{\vec I}_n} \cdot {{\hat C}_{nm}}(\left| {{S_z}} \right|) \cdot {{\vec I}_m}} \text{.}
\end{aligned}
\end{equation}
The relatively large zero-field splitting does not allow the electron spin to flip and thus we can
make the so-called secular approximation, removing all terms which allow direct electronic spin flips. Therefore, we can reduce the Hilbert space of the system by projecting Hamiltonian onto each of the electron spin states. We can rewrite the projected
Hamiltonian, ${P_{{m_s}}}H{P_{{m_s}}}$ (where ${P_{{m_s}}} = \left| {{m_s}} \right\rangle \left\langle {{m_s}} \right|$) as:
\begin{equation}
\begin{aligned}
{H_{{m_s}}} = &D\left| {{m_s}} \right| - {\gamma _e}{B_z}{m_s} + \sum\limits_n {\vec \Omega _n^{({m_s})} \cdot {{\vec I}_n}} \\
&+ \sum\limits_{n > m} {{{\vec I}_n} \cdot \hat C_{nm}^{{m_s}}(\left| {{S_z}} \right|) \cdot {{\vec I}_m}} \text{.}
\end{aligned}
\end{equation}
The projected Hamiltonian under a rotating-wave frame of MW is
\begin{equation}
\begin{aligned}
H = \Delta  + \sum\limits_n {\vec \Omega _n^{({m_s})} \cdot {{\vec I}_n}}  + \sum\limits_{n > m} {{{\vec I}_n} \cdot \hat C_{nm}^{{m_s}}(\left| {{S_z}} \right|) \cdot {{\vec I}_m}} \text{,}
\end{aligned}
\end{equation}
where $m_s$ denotes the electron spin state, $\vec \Omega _n^{{m_s}}$ is the effective Larmor vector for nucleus $n$, and $C_{nm}$ is the effective coupling between nuclei $n$ and $m$.
Therefore, the signal can be expressed as:
$${s_\Delta } = Tr\left[ {\frac{{\left( {I + {\sigma _x}} \right)}}{2}{U^{\frac{N}{2}}}\frac{{\left( {I + {\sigma _x}} \right)}}{2} \otimes {\rho _d}{U^{\dag \frac{N}{2}}}} \right] \text{,}$$
$$U = {e^{ - i{H_1}\tau /\hbar }}R(\pi ){e^{ - i{H_2}\tau /\hbar }}{e^{ - i{H_2}\tau /\hbar }}R(\pi ){e^{ - i{H_1}\tau /\hbar }}\text{,}$$
$${H_1} = \Delta  + \sum\limits_n {\vec \Omega _n^{({m_s})} \cdot {{\vec I}_n}}  + \sum\limits_{n > m} {{{\vec I}_n} \cdot \hat C_{nm}^{{m_s}}(\left| {{S_z}} \right|) \cdot {{\vec I}_m}}\text{,} $$
$${H_2} = \Delta + \sum\limits_n {\vec \Omega _n^{({m_s})} \cdot {{\vec I}_n}}  + \sum\limits_{n > m} {{{\vec I}_n} \cdot \hat C_{nm}^{{m_s}}(\left| {{S_z}} \right|) \cdot {{\vec I}_m}}\text{,} $$
where $\rho  = {\rho _{NV}} \otimes {\rho _d}$ is the probe state, ${\rho _{NV}} = \left| 0 \right\rangle \left\langle 0 \right|$ and ${\rho _d} = {\left( {I/2} \right)^{ \otimes n}}$ is the maximally mixed state of dark nuclear spins under high temperature approximation.

We can always parameterize $R(\pi)$ with Pauli matrix of spin $1/2$ in this way:
\begin{equation}
\begin{aligned}
R(\pi ) =& {a_0}I + {a_x}{\sigma _x} + {a_y}{\sigma _y} + {a_z}{\sigma _z}\\
 =& {a_0}I + \sqrt {a_x^2 + a_y^2} {\sigma _n} + {a_z}{\sigma _z}\\
 =& {a_0}I + \sqrt {{F_{QS}}} {\sigma _n} + {a_z}{\sigma _z}\text{,}
\end{aligned}
\end{equation}
where ${\sigma _n} = \left( {\frac{{{a_x}}}{{\sqrt {{F_{QS}}} }},\frac{{{a_y}}}{{\sqrt {{F_{QS}}} }},0} \right) \cdot \overrightarrow \sigma  $.  Due to the orthogonal normalization of generator, we will get ${s_\Delta }(\Delta ) = F_{QS}^N{s_{\Delta  = 0}} + e\left( {{{\left( {\frac{\Delta }{\Omega }} \right)}^2}} \right)$, where $N$ denotes the number of $\pi$ pulses of DD sequences. For NV center ensemble, if the unit operation of DD method is failed, the inhomogeneous broadening effect will dominate the decoherence of system and the corresponding detection signal component decreases with time scaled as $T_2^{\text{*}}$  \cite{Sxu2018room}. Once ${\omega _s}T_2^* > 1$, those components ($e\left( {{({ {\frac{\Delta }{\Omega }}})^2}} \right)$) can be neglected and the intensity of signal is scaled as $F_{QS}^N$.


\end{document}